\documentstyle[floats,prd,aps,epsf,eqsecnum,12pt]{revtex}
\makeatletter
\newbox\tempboxa
\newdimen\captionboxsubcount 
\def\capsize#1{\captionboxsubcount=#1pt}
\newdimen\captionboxsub
\captionboxsub=\hsize \advance\captionboxsub by -\captionboxsubcount
\advance\captionboxsub by -\captionboxsubcount
\long\def\@makecaption#1#2{
 \setbox\@tempboxa\hbox{#1 #2}
 \ifdim \wd\@tempboxa >\captionboxsub 
\rightskip=\captionboxsubcount \leftskip=\captionboxsubcount #1 #2 
\else \hbox to\hsize{\hfil\box\@tempboxa\hfil} 
 \fi}
\makeatother
\capsize{30}

\begin{document}

\begin{titlepage}

\begin{flushright}
\begin{minipage}{5cm}
\begin{flushleft}
\small
\baselineskip = 13pt
YCTP-P28-97\\
SU--4240--673\\
hep-th/9801097 \\
\end{flushleft}
\end{minipage}
\end{flushright}

\begin{center}
\Large\bf
Anomaly induced QCD potential and Quark Decoupling
\end{center}

\vfil

\footnotesep = 12pt

\begin{center}
\large
Stephen {\sc D.H.~Hsu } 
\footnote{Electronic address: {\tt hsu@hsunext.physics.yale.edu}}
\quad \quad
Francesco {\sc Sannino}
\footnote{
Electronic address : {\tt sannino@genesis1.physics.yale.edu}}\\
{ \it  \qquad Department of Physics, Yale University, New Haven, CT 06520-8120, USA.}\\
\vskip .5cm
Joseph {\sc Schechter}
\footnote{
Electronic address : {\tt schechte@suhep.phy.syr.edu}}\\
{\it 
\qquad
Department of Physics, Syracuse University, 
Syracuse, NY 13244-1130, USA.}
\end{center}

\vfill
\begin{center}
\bf
Abstract
\end{center}
\begin{abstract}
\baselineskip = 17pt
We explore the anomaly induced effective QCD meson potential in the framework of the effective Lagrangian 
approach. We suggest a decoupling procedure, when a flavored quark becomes massive,  
which mimics the one employed by Seiberg for supersymmetric gauge theories. It is seen 
that, after decoupling, the QCD potential naturally converts to the one with one less flavor. 
We study the $N_c$ and $N_f$ dependence of the $\eta^{\prime}$ mass.
\end{abstract}
\begin{flushleft}
\end{flushleft}

\vfill

\end{titlepage}

\setcounter{footnote}{0}

\section{Introduction}
In the last few years there has been a great flurry of interest in the 
effective Lagrangian approach to supersymmetric gauge theories. This was 
stimulated by some papers of Seiberg \cite{Seiberg} and Seiberg and Witten 
\cite{Seiberg-Witten} in which a number of fascinating ``exact results'' were 
obtained. 
In particular Seiberg has provided a very interesting picture for different 
phases of supersymmetric gauge theories. 
There are already several interesting review articles 
\cite{IntSeiberg,Peskin,DiVecchia}. 

It is natural to hope that information obtained from the more highly 
constrained  supersymmetric gauge theories can be used to learn more 
about ordinary gauge theories, notably QCD.  
In a recent paper \cite{toy} it was shown how the effective Lagrangian for super QCD 
might go over to the one for ordinary QCD by a mechanism whereby the gluinos and squarks 
in the fundamental theory decouple below a given supersymmetric breaking scale $m$. 
To accomplish this goal a suitable choice of possible supersymmetry breaking terms 
was used. An amusing feature of the model was the emergence 
of the ordinary QCD degrees of freedom which had been hidden in the auxiliary fields of 
 the supersymmetric effective Lagrangian. 
Constraints on the supersymmetry breaking 
terms were obtained by requiring the trace of the energy momentum tensor to agree 
(at one loop level), 
once supersymmetry  was broken,  with that of ordinary QCD.  
It was also noticed that a reasonable initial picture resulted by neglecting the K\"ahler terms in 
the original supersymmetric effective Lagrangian. This feature is analogous to Seiberg's 
treatment  \cite{Seiberg} of supersymmetric effective Lagrangians with different numbers of flavors. It led to a dominant piece of the QCD effective Lagrangian possessing a kind 
of tree level "holomorphicity". Physically this corresponds to the explicit realization 
of the axial and trace anomalies by the model. 

In this letter we will further explore the "holomorphic" part of the potential for QCD with 
$N_f(<N_c)$ massless quarks as obtained by breaking super QCD according to the scheme above (see section III of \cite{toy}). In particular, we shall study the analog of the decoupling procedure used in \cite{Seiberg} when one flavor becomes heavy. In the present case we no longer have the protection of supersymmetry for deriving exact results so the procedure will be carried out at the one loop level. The analog of the Affleck-Dine-Seiberg  (ADS) superpotential \cite{refADS} turns out to be \cite{toy} the holomorphic part of the potential:
\begin{equation}
V(M)=-C\left(N_c,N_f\right) 
\left[\frac{\Lambda^{\frac{11}{3}N_c - \frac{2}{3} N_f}}{{\rm det} M}\right]^
{\frac{12}{11\left(N_c - N_f\right)}} + {\rm h.c.} \  ,
\label{adsqcd}
\end{equation}
where $\Lambda$ is the invariant QCD scale and the meson matrix $M$ contains $N_f^2$ scalars and $N_f^2$ pseudoscalars. A particular choice of the dimensionless positive quantity $C\left(N_c,N_f\right)$ was made in \cite{toy}.  
Here we find the constraints on $C\left(N_c,N_f\right)$ which follow from decoupling. The result can be used to suggest that the well known \cite{Wittena,Veneziano} large $N_c$ behavior of the $\eta^{\prime}$ (pseudoscalar singlet) meson mass should also include an $N_f$ dependence of the form:
\begin{equation}
M^2_{\eta^{\prime}}\propto\frac{N_f}{N_c - N_f} \Lambda^2 \qquad \left(N_f < N_c\right)\ .
\end{equation}
This could be considered an indication that, while $M_{\eta^{\prime}}$ is suppressed in the large $N_c$ limit, there is an enhancement mechanism for $N_f$ near $N_c-1$ to explain its relatively large numerical value.

\section{Meson Potential and Quark Decoupling}
The various super QCD (SQCD) effective Lagrangians \cite{refADS} are essentially derived using the anomaly structure of the underlying gauge theory as well as supersymmetry. It is thus interesting to first see how much of the effective QCD Lagrangians we get after decoupling the super-partner fields, follows just from anomaly considerations. In ordinary QCD the two relevant anomalies are the trace and Adler-Bell-Jackiw $U_A(1)$ anomalies:
\begin{eqnarray}
\theta^m_m&=&-b\frac{g^2}{32 \pi^2} F^{mn}_aF_{mn;a}\equiv 2 b H \ , \\
\delta_{U_A(1)}{ V_{QCD}}&=&N_f \alpha \frac{g^2}{32\pi^2} 
\epsilon_{mnrs}F^{mn}_aF^{rs}_a\equiv 4 N_f \alpha G \ .
\label{anomalies}
\end{eqnarray}
$N_f$ is the number of flavors,  $\displaystyle{b=\frac{11}{3}N_c - \frac{2}{3}N_f}$ is the coefficient of the one loop beta function and left handed quarks have unit axial charge. $H$ and $G$ can be interpreted as composite operators describing, in the confining regime, the scalar and pseudoscalar glueball fields \cite{joe}.  
The general effective (i.e. in terms of composite operators) potential which, at tree level,  saturates  the one loop anomalies is \cite{joe}:
\begin{equation}
V=-b{F}\sum_n \frac{c_n}{n} {\rm ln} \left(\frac{{\cal O}_n}{\Lambda^n}\right) + h.c. +  {V_I}  \ ,
\label{general}
\end{equation} 
where ${\cal O}_n$ is an operator built out of the relevant degrees of freedom at a given scale $\mu$ (glueballs, mesons, baryons, ..)  with mass dimension $n$ and axial charge $q_n$,  ${F}=H+i\delta G$ and $V_I$ is a scale and $U_A(1)$ invariant potential. 
The anomaly constraints are: 
\begin{equation}
\sum_n c_n =1 \ , \qquad
\sum_n \frac{c_n}{n} q_n =\frac{2N_f }{b \delta} .
\label{sum}
\end{equation}
 $\Lambda$ is connected with the invariant scale of the theory and the $\theta$ parameter 
(at  one loop) via $\displaystyle{\Lambda^{b}=\mu^b e^{-\tau}}$ with $\tau=\frac{8\pi^2}{g^2(\mu)} - i\theta$.
At high scales ( $\mu \gg\Lambda$) we expect to approach the classical regime where the glue potential  
displays the holomorphic structure $-\tau (H+iG) + h.c.$. This suggests choosing $\delta=1$ and   hence $F=H+iG$. 
Here we assume that the lightest meson  ($M$)\footnote{In Ref.~{\cite{toy}} 
$M\simeq- \bar{q}_R  q_L$ was denoted $F_T$.} and glueball ($H$ and $G$) fields are the relevant degrees of freedom. 
It is amusing to notice that the field associated with the $U_A(1)$ transformation ($\eta^{\prime}$) can only appear in the first term of  the potential in Eq.~(\ref{general}). 

We can arrange the anomaly potential (first term in Eq.~(\ref{general})) to be "holomorphic"  
by setting ${\cal O}_n={\cal O}_n\left(F,{\rm det}M\right)$. 
By holomorphic we mean a potential of the form $\chi(\Phi) + h.c.$ where $\chi$ is a function of the generic complex field $\Phi$.  
Now, in Ref.~{\cite{toy}} it was shown that a suitable decoupling of the holomorphic superpotential in the Taylor Veneziano Yankielowicz (TVY)  model \cite{refADS} yielded an important "holomorphic" contribution to the ordinary potential for QCD. {}For $N_f(<N_c)$ flavors the result was of the form
\begin{eqnarray}
V(F,M)&=&-\frac{11}{12}\left(N_c - N_f\right) F \left[{\rm ln}\left(
\frac{-N_c F}{\gamma \Lambda^4}\right) -1\right] - F{\rm ln}\left(\frac{{\rm det} M}{\Lambda^{3N_f}}\right) \nonumber \\
&&+A\left(N_f,N_c\right) F + {\rm h.c.} \ ,
\label{tvyqcd}
\end{eqnarray}
where $\displaystyle{\gamma=\frac{12N_c - 4N_f}{11N_c - 2N_f}}$ and $A\left(N_c , N_f \right)$
is a dimensionless constant which cannot be fixed by saturating the QCD anomalies and  assuming holomorphicity. This is because the term $AF$ does not contribute to $\theta^m_m$ and is also a chiral singlet. 
The potential in Eq.~(\ref{tvyqcd}) is easily seen to be consistent with the general form in Eq.~(\ref{general}), and the associated constraints in Eq.~(\ref{sum}) when the scale dimension of $M$ is fixed to be $3$ and $\delta=1$.  (The constants are 
$\displaystyle{c_4=1-3\frac{N_f}{b}} $ and $\displaystyle{c_{3N_f}=3\frac{N_f}{b}}$ ). 

By integrating out the "heavy" glueball field $F$  one obtains the "light" meson field potential 
for $M$. This is given in Eq.~(\ref{adsqcd}) wherein  the unknown coefficient $C\left(N_c,N_f\right)$ is related to $A(N_c,N_f)$ by the 
following expression:
\begin{equation}
A\left(N_c,N_f\right)=\frac{11}{12}\left(N_c - N_f\right) {\rm ln} \left(\frac{12}{11}\frac{N_c}{\gamma\left(N_c - N_f \right)} 
C\left(N_c,N_f\right)\right) \ .
\label{AC}
\end{equation}
Similar effective potential models and some related phenomenological questions have already been 
discussed in the literature \cite{MS,SST,GJJS,GJJS2,smrst}.

The potential for the meson variables in Eq.~(\ref{adsqcd}) is similar to the effective  
ADS superpotential for massless super QCD theory with $N_f<N_c$  
\cite{refADS}
\begin{equation}
W_{ADS}(T)=-\left(N_c-N_f\right) 
\left[\frac{\Lambda^{3N_c -N_f}_S}{{\rm det} T}\right]^{\frac{1}{N_c-N_f}} \ ,
\label{ads}
\end{equation}
where $T \simeq Q\tilde{Q}$ is 
the composite meson superfield, $Q$ and $\tilde{Q}$ are the quark super fields and 
$\Lambda_S$ is the invariant scale of SQCD. In the instanton generated 
super potential the exponent of $\Lambda_S$ is the coefficient of the super 
symmetric beta function. 
In the potential for the ordinary meson variables Eq.~(\ref{adsqcd})
 the exponent inside the square brackets of the QCD invariant scale 
$\Lambda$ is similarly the coefficient 
of the one loop QCD beta function 
$\displaystyle{b=\frac{11}{3}N_c - \frac{2}{3}N_f}$.   
As in  the SQCD case the ordinary QCD potential Eq.~(\ref{adsqcd}) 
can also be constructed if we assign to $\Lambda^{b}$ charge $2N_f$ under 
the matter field axial transformation, as prescribed by the $\theta$-angle variation, and 
if holomorphicity in the coupling constant is assumed.  

An important difference with respect to the ADS superpotential, as already explained in 
Ref.~{\cite{toy}}, is the fact that the QCD potential displays a fall to the origin 
while the ADS potential does not have a minimum for finite values of the squark condensate. 
In \cite{toy} it was shown that the fall to the origin 
can be cured by introducing non holomorphic 
terms which also trigger spontaneous chiral symmetry breaking.

The mesonic holomorphic potential, due to the presence of the appropriate beta function, 
is expected to hold  for any $N_f<N_c$. In particular a non trivial check 
would be 
to decouple a single heavy flavor and show that the potential reduces to the potential 
for one less flavor fermion. A similar procedure, at the supersymmetric level, was used by 
Seiberg to check the validity of the ADS superpotential. This is in the same spirit as the well 
known \cite{Wittendec} criterion for decoupling a heavy flavor (at the one loop level). 
{}For a small mass $m$ of the $N_f $ -th quark the perturbative 
 contribution, prescribed in the fundamental lagrangian, 
of the mass operator to the potential is 
\begin{equation} 
V(M) = {\cdots} -mM^{N_f}_{N_f}  + {\rm h.c.} \ .
\end{equation}
To achieve a complete decoupling we generalize the perturbative 
mass operator to  
\begin{equation}
V_{m}=-m^{\Delta}{M^{N_f}_{N_f}}^{\Gamma}  + {\rm h.c.}\ , 
\label{breaking}
\end{equation}
where dimensional analysis requires $\Delta=4-3\Gamma$. We
 interpret the departure from unity of $\Delta$ as an 
effective  dynamical evolution of the chiral symmetry 
breaking operator for large values of $m$.    
The total potential for large $m$, 
\begin{equation}
V_{T}(M)=-C\left(N_c,N_f\right) 
\left[\frac{\Lambda^{\frac{11}{3}N_c - \frac{2}{3} N_f}}{{\rm det} M}\right]^
{\frac{12}{11\left(N_c - N_f\right)}} -m^{\Delta}{M^{N_f}_{N_f}}^{\Gamma}  + {\rm h.c.} \ ,
\label{VT}
\end{equation}
where $\Lambda$ is the appropriate invariant scale for $N_f$ flavors and $N_c$ colors, 
must convert to the potential 
for $N_f-1$ light quarks.  
Notice that the generalized mass term preserves the holomorphic structure. Here it is being assumed that the non-holomorphic terms in the potential can be treated as small pertubations. 
A similar mass 
operator was already introduced in Ref.~{\cite{toy}}  to appropriately decouple the gluino in 
the underlying theory. 
Eliminating the heavy degrees of freedom by their equations of motion  and substituting back in the potential gives
\begin{equation}
V(\hat{M})=-\left[\frac{\Gamma + w}{w}\right] \, \left[\frac{C\left(N_c,N_f\right)w} 
{\Gamma}\right]^{\frac{\Gamma}{\Gamma + w}}
\left[\frac{\Lambda^{b(N_c,N_f)} m^{\frac{\Delta}{\Gamma}}}{{\rm det}\hat{M}}\right]
^{\frac{\Gamma w}{\Gamma + w}}  + {\rm h.c.}\ ,
\label{preqcd}
\end{equation}
where $\displaystyle{w=\frac{12}{11\left(N_c-N_f\right)}}$, $b(N_c,N_f)$ is the coefficient 
of the beta function for the theory with $N_c$ colors and $N_f$ flavors and 
$\hat{M}$ is the meson matrix for $N_f-1$ flavors.  
In the standard physical picture the gauge coupling constant evolves according to the 
QCD beta-function for $N_f$ flavors above scale $m$ and according to the QCD beta-function 
for $N_f-1$ flavors below scale $m$. Since the coupling constant at scale $\mu$ is given 
by $\displaystyle{\left(\frac{\Lambda}{\mu}\right)^b= {\rm exp}\left(-\frac{8\pi^2}{g^2(\mu)}\right)}$, the matching at $\mu =m$ requires 
$\displaystyle{\left(\frac{\Lambda_{N_c,N_f}}{m}\right)^{b\left(N_c,N_f\right)}=
\left(\frac{\Lambda_{N_c,N_f-1}}{m}\right)^{b\left(N_c,N_f-1\right)}}$, which yields
\begin{equation}
\Lambda_{N_c,N_f-1}^{b\left(N_c,N_f-1\right)}=\Lambda_{N_c,N_f}^{b\left(N_c,N_f\right)} 
m^{\frac{2}{3}} \ .
\label{match}
\end{equation}
If we require the presence of the chiral symmetry breaking term in Eq.~(\ref{VT}) to convert 
the $N_f$ theory into the $N_f -1$ theory,  
the ratio $\displaystyle{\Delta/\Gamma}$ is fixed  to $\displaystyle{\frac{2}{3}}$ and by using 
the relation $\Delta=4-3\Gamma$ one derives 
\begin{equation} 
\Delta=\frac{8}{11}, \qquad \Gamma=\frac{12}{11} \ .
\label{DG}
\end{equation}  
This value\footnote{These are the same exponents found in section II of \cite{toy} for decoupling a massive gluino. Here we can also use, as  in \cite{toy},  the one loop matching of the 
anomaly of the traced energy momentum tensor to derive the same exponents.} of $\Gamma$ is exactly what is needed to get the correct exponent  
\begin{equation}
\frac{\Gamma w}{\Gamma + w} =\frac{12}{11\left(N_c - N_f +1\right)} \ ,
\end{equation}
in Eq.~(\ref{preqcd}).  
A consistent decoupling is obtained provided that the coefficient $C\left(N_c,N_f\right)$ obeys the following recursive relation
\begin{equation}
\left[\frac{C\left(N_c,N_f\right)}{N_c - N_f}\right]^{N_c-N_f} 
=\left[\frac{C\left(N_c,N_f-1\right)}{N_c - N_f+1}\right]^{N_c-N_f+1} \ . 
\label{recursive}
\end{equation}
The general solution of Eq.~(\ref{recursive}) is finally, 
\begin{equation}
C\left(N_c,N_f\right)=\left(N_c - N_f\right) D\left(N_c\right)^{\frac{1}{N_c-N_f}} \ .
\label{linear}
\end{equation}
Using Eq.~(\ref{AC}), $A\left(N_c,N_f\right)$ can 
be expressed as the function of ${D}$: 
\begin{equation} 
A\left(N_c,N_f\right)=\frac{11}{12} \left(N_c-N_f\right) {\rm ln}\left(\frac{12}{11}\frac{N_c}{\gamma}\right) + \frac{11}{12} {\rm ln}{D}\left(N_c\right) \ .
\label{linearb}
\end{equation}

It is interesting to contrast the result in Eq.~(\ref{linear}) for the coefficient of the "holomorphic" part of the QCD potential Eq.~(\ref{adsqcd}) with Seiberg's result \cite{Seiberg} $C\left(N_c,N_f \right)=\left(N_c-N_f\right)$ for the coefficient of the ADS superpotential in Eq.~(\ref{ads}). Clearly, in the SUSY case the analog of $D(N_c)$ is just a constant (which turns out, in fact, to be unity by an instanton analysis \cite{Seiberg}). This feature arises in the SUSY case because of the existence of squark fields which can break the gauge and flavor symmetries by the Higgs mechanism; it leads to $C\left(N_c,N_f\right)=C\left(N_c - N_f\right)$. Since there are no appropriate scalar fields in ordinary QCD we do not expect such a feature. The possibility of a non-constant $D(N_c)$ factor can thus be taken as an indication that there is no Higgs mechanism present.

\section{Physical Effects}
As mentioned above, we must add to the "holomorphic" piece of the potential (Eq.~(\ref{adsqcd}) with Eq.~(\ref{AC})) a "non-holomorphic" piece in order to prevent a "fall to the origin". This would also yield a non-zero value for $\langle M^b_a\rangle$. The non-holomorphic piece will be required to neither contribute to the trace anomaly nor to the axial anomaly. At the level of Eq.~(\ref{general}), where the "heavy" glueball fields  have not yet been integrated out, such a potential could be of the form
\begin{equation}
V_I =\sum_n A_n \left(F^{*}F\right)^{\frac{1}{2} -\frac{3}{4}n} \, {\rm Tr} 
\left[\left(M^{\dagger}M\right)^n\right] \ ,
\label{nholo}
\end{equation}
where the $A$'s are real constants. (For definiteness the terms have been restricted to those which are leading in the large $N_c$ limit). For our initial analysis we assume that $V_I$ is small compared to the holomorphic piece and can be treated as a pertubation. An example of a single 
term with $n>\frac{2}{3}$ was presented in Ref.~{\cite{toy}}. 
An analogous term without the heavy $F$ fields is 
\begin{equation}
V^{\prime}_I=\sum_n A^{\prime}_n {\rm Tr} \left[\left(M^{\dagger}M\right)^n\right] \ ,
\label{nholo2}
\end{equation}
which should be added to Eq.~(\ref{adsqcd}). The new feature of the present model is that the piece which mocks up the anomalies can be thought of as inheriting a specific holomorphic structure from the SQCD theory. 

Since the model of Eq.~(\ref{adsqcd}) plus Eq.~(\ref{nholo2}) contains a great deal of arbitrariness it is natural to wonder what can be learned from it. The key point is that Eq.~(\ref{nholo2}) possesses an axial $U\left(N_f\right)_L \times U\left(N_f\right)_R$ symmetry and cannot contribute to the $\eta^{\prime}$ (pseudoscalar singlet) mass. Thus we may learn something about this quantity from Eq.~(\ref{adsqcd}) directly. Now the $\eta^{\prime}$ field may be isolated from the meson matrix $M$ as
\begin{equation}
M=K\tilde{U}\exp\left[i\frac{\eta^{\prime}}{\alpha \Lambda \sqrt{N_f}}\right] \ , 
\label{dec}
\end{equation}
where $K=K^{\dagger}$, $\tilde{U}^{\dagger}=\tilde{U}^{-1}$, ${\rm det}\tilde{U}=1$ and  
$\alpha$ is a dimensionless real constant. $K$ contains the scalar fields and $\tilde{U}$ the non flavor-singlet pseudoscalars. For our present purpose we may set $K=\beta \Lambda^3 {\bf 1}$, where $\beta$ is another dimensionless real constant. For simplicity we shall make the approximation $\alpha=\beta=1$, which will not affect our results in the large $N_c$ limit. Then substituting Eq.~(\ref{dec}) into Eq.~(\ref{adsqcd}), using
Eq.~(\ref{linear}) and expanding the result to quadratic order in $\eta^{\prime}$ yields the mass
\begin{equation}
M^2_{\eta^{\prime}} =\frac{2N_f}{N_c - N_f}\left(\frac{12}{11}\right)^2 \Lambda^2 
D\left(N_c\right)^\frac{1}{N_c - N_f} \ .
\label{etamass}
\end{equation}
Here it has been assumed that the $\eta^{\prime}$ has the canonical Lagrangian mass term 
$\displaystyle{-\frac{1}{2} \partial_m \eta^{\prime} \partial^m  \eta^{\prime}}$.  If the factors $\alpha$ and $\beta$ were to be included the right hand side of Eq.~(\ref{etamass}) would get 
multiplied by $\displaystyle{\alpha^{-2} \beta^{-\frac{12N_f}{11\left(N_c -N_f\right)}}}$. 
$\alpha$ does not have any large $N_c$ dependence and that of $\beta$ gets washed out. 

$D(N_c)$ is unspecified by our model so we cannot predict $M_{\eta^{\prime}}$. We can however check the dependence on $N_c$ of the quantities in Eq.~(\ref{etamass}) for large $N_c$. It is well known  \cite{Wittena,Veneziano} that 
$\displaystyle{M^2_{\eta^{\prime}}\sim 1/N_c}$ for large $N_c$. Furthermore, $\langle F\rangle$ (defined after Eq.~(\ref{general})) should behave like $N_c$ for large $N_c$ by standard counting arguments. Hence (see the discussion after Eq.~(2.18) of Ref.~\cite{toy}) $\Lambda$ is expected to behave as 
$\displaystyle{\left(N_c\langle F\rangle\right)^{\frac{1}{4}}\sim N_c^{\frac{1}{2}}}$ at large $N_c$. For consistency we thus expect the large $N_c$ behavior
\begin{equation}   
D\left(N_c\right) \sim \left(\frac{1}{N_c}\right)^{N_c} \ .
\label{consistence}
\end{equation}
It is amusing to observe that 
when $N_f$ is close to $N_c$ the resulting pole in Eq.~(\ref{etamass}) suggests a possible 
enhancement mechanism for the $\eta^{\prime}$ mass. This could explain the unusually 
large value of this quantity in the realistic three flavor case.

Of course the $N_f=N_c$ case is rather non-trivial. Equation (\ref{adsqcd}) shows that the overall coefficient of our model, $C\left( N_c, N_f \right)$ should vanish for $N_f=N_c$. This is what happens in the SQCD situation. There, Seiberg \cite{Seiberg} shows that the effective superpotential should be replaced by a quantum constraint which also involves the baryon superfields. Before considering the analog of these results in our context, let us try to understand the situation a little better by going back to the holomorphic part of the potential in Eq.~(\ref{tvyqcd}), in which the heavy glueball field $F$ has not yet been integrated out. 
Focusing on the pieces of  Eq.~(\ref{tvyqcd}) which are related to generating an 
$\eta^{\prime}$ mass we get in a {\it schematic} form 
\begin{equation}
V\left(F,M\right) \sim - \left(N_c - N_f\right) G^2 + \eta^{\prime} G \ ,
\label{apx}
\end{equation}
where $G$  is the pseudoscalar glueball field defined in Eq.~(\ref{anomalies}) and inessential factors have been dropped. The first term is a wrong sign (for $N_f < N_c$) mass term for $G$ and is obtained by expanding the first term of Eq.~(\ref{tvyqcd}). Integrating out $G$ according to the equation of motion results in a correct sign mass term for the $\eta^{\prime}$ as long as 
$N_f < N_c$. It is clear that this mechanism will not work for $N_f \ge N_c$ From our present treatment we can see what is likely to be happening in the case of ordinary QCD. Here the non-holomorphic piece in Eq.~(\ref{nholo}) may contain a $-G^2$ piece when it is expanded and may thus generate an $\eta^{\prime}$ mass. We have been treating $V_I$ in Eq.~(\ref{nholo}) 
as a small pertubation but, for consistency, it must be non-negligible especially for $N_f\ge N_c$. The usual phenomenology is consistent with the large $N_c$ approximation to QCD so perhaps it is easiest to use the present model in that case too. In that event we would always have $N_f \ll N_c$. However it is certainly of interest to examine the small $N_c$ case. The present model suggests that something different is occurring as we let $N_f$ vary near $N_c$. It is noteworthy that recent preliminary computer simulations \cite{mawhinney} indicate that there is a rather noticeable change in the quark condensate for $N_f$ near $N_c$. 

Now let us return  to the discussion of the holomorphic part of the potential and point out that formally the $N_f=N_c$ case can be handled as in SQCD. When we put $N_f=N_c$ in Eq.~(\ref{tvyqcd}) 
\begin{equation}
V\left(F,M\right) \rightarrow - F {\rm ln}\, \left(\frac{{\rm det}M}
{e^{A\left(N_c,N_c\right)} \Lambda^{3 N_c}_{N_c,N_c}}\right) + h.c. \ .
\end{equation} 
Integrating out $F$ then gives the {\it constraint} on the field $M$, 
\begin{equation}
{\rm det} M = e^{A(N_c,N_c) }\Lambda^{3N_c} _{N_c,N_c} \ ,
\label{constraint}
\end{equation}
rather than a potential  $V\left(M\right)$. In fact the holomorphic potential vanishes as we argued. Nevertheless this vanishing potential is consistent (as in the SQCD case) with decoupling to the $N_f=N_c -1 $ theory. Following Eq.~(\ref{breaking}) we consider the $N_c=N_f$ potential with a heavy $N_f$-th quark of mass $m$:
\begin{equation}
V=0 - m^{\Delta} \left(M^{N_f}_{N_f}\right)^{\Gamma} + h.c. \ ,
\label{null}
\end{equation}
where $\Delta=8/11$ and $\Gamma=12/11$ as before. Now the constraint in Eq.~(\ref{constraint}) enables us to write 
\begin{equation}
M_{N_f}^{N_f} {\rm det} \hat{M} \rightarrow e^{A\left(N_c,N_c\right)} \Lambda^{3N_c}_{N_c,N_c} \ ,
\end{equation}
where $\hat{M}$ is the meson matrix for $N_c-1$ flavors. Substituting this back into Eq.~(\ref{null}) yields 
\begin{equation}
V=-e^{A\left(N_c,N_c\right) \Gamma} \left(
\frac{m^{\frac{\Delta}{\Gamma}} \Lambda^{3N_c}_{N_c,N_c}} { {\rm det} \hat{M} }\right)^{\Gamma}  + h.c. \ . 
\end{equation}
When one uses Eq.~(\ref{match}) and Eq.~(\ref{linearb}) this is recognized as precisely Eq.~(\ref{adsqcd}) together with Eq.~({\ref{linear}})  for the $N_f=N_c -1$ case. This argument may also serve to clarify what is "happening" in the SQCD situation. 

A final question concerns the role of baryons when $N_f=N_c$. Baryon superfields  were shown 
to play an important role \cite{Seiberg} for $N_f=N_c$ in SQCD. One might hope to decouple the superpartners of the baryon superfields $B$ and $\tilde{B}$  \cite{Seiberg} to obtain information on the QCD three quark baryons. However it is easy to see that none of the components of $B$ and $\tilde{B}$ contain any three quark composite fields. Thus the quick answer would be that the baryon fields are irrelevant for the ordinary QCD effective Lagrangian. This point of view is perhaps reinforced by the usual treatment (for large $N_c$) of baryons as solitons of the effective meson Lagrangian. Nevertheless, the arguments in the first part of section II show that ordinary baryon fields may readily be added to the holomorphic potential in a manner consistent with the anomaly structure of QCD. They may very well be relevant for a small $N_c$ treatment of the theory. We hope to return to this question elsewhere.

 \acknowledgments
One of us (F.S.) is happy to thank Thomas Appelquist and Noriaki Kitazawa for interesting discussions. We would also like to thank Amir Fariborz for helpful comments on the manuscript.
The work of S.H. and F.S. has been partially supported by the US DOE  under contract 
DE-FG-02-92ER-40704. 
The work of J.S. has been supported in part by the US DOE under contract
DE-FG-02-85ER 40231.

\end{document}